\documentclass[12pt,preprint]{aastex}
\received{2004 May 12}
\begin{document}
\title{The First Measurements of Galaxy Clustering from IRAC Data
of the Spitzer First Look Survey }
\author{Fan Fang, David L. Shupe, Gillian Wilson, Mark Lacy, Dario Fadda, Tom Jarrett\altaffilmark{1}, Frank Masci, P. N. Appleton, Lee Armus, Scott Chapman, Philip I. Choi, D. T. Frayer, Ingolf Heinrichsen, George Helou, Myungshin Im\altaffilmark{2}, Francine R. Marleau, B. T. Soifer, Gordon K. Squires, L. J. Storrie-Lombardi, Jason Surace, Harry I. Teplitz, Lin Yan}
\affil{Spitzer Science Center, Caltech 314-6, Pasadena, CA 91125}
\altaffiltext{1}{Infrared Processing and Analysis Center, Caltech 100-22, Pasadena, CA 91125} 
\altaffiltext{2}{School of Earth and Environmental Sciences, Seoul National University, Seoul, S. Korea}

\begin{abstract}
We present the first results of the angular auto-correlation function
of the galaxies detected by the Infrared Array Camera (IRAC) instrument
in the First Look Survey (FLS) of the Spitzer Space Telescope.  We detect
significant signals of galaxy clustering within the survey area.
The angular auto-correlation function of the galaxies detected in each
of the four IRAC instrument channels is consistent with a power-law
form $w(\theta)=A\theta^{1-\gamma}$ out to $\theta = 0.2\arcdeg$,
with the slope ranging from $\gamma = 1.5$ to $1.8$.  We estimate
the correlation amplitudes $A$ to be $2.95 \times 10^{-3}$,
$2.03 \times 10^{-3}$, $4.53 \times 10^{-3}$,
and $2.34 \times 10^{-3}$ at $\theta=1\arcdeg$ for galaxies detected
in the IRAC 3.6$\mu$m, 4.5$\mu$m, 5.8$\mu$m, and 8.0$\mu$m instrument
channels, respectively.  We compare our measurements at 3.6$\mu$m with
the previous K-band measurements, and discuss the implications of these
results.
\end{abstract}


\section{Introduction}
\label{sec:intro}

Since the early photographic-plate surveys the angular 2-point correlation
function has been used to first characterize the galaxy spatial distribution
in a catalog \citep{gp77}.
Combined with selection function models in the third dimension these
characterizations can yield information of the galaxy distribution in
space and time, providing important clues to how structures form
and evolve.

So far these statistical characterizations have been established for galaxies
in wavelengths that are systematically probed to various sensitivity levels.
At the near-infrared K-band, for example, the angular galaxy clustering has
been extensively measured to various depths
\citep{baugh96,roche98,roche99,roche02,roche03,daddi00,kw00,mc00,maller03}.
It is interesting to compare these K-band measurements with the ones
from the Infrared Array Camera (IRAC) onboard the Spitzer Space telescope,
since they detect the same population of evolved star-forming systems, as
the rest-frame K-wavelength shifts into the IRAC wavelengths at higher redshifts.

The First Look Survey (FLS) using IRAC opens up new windows in near to
mid-infrared wavelengths and detects galaxies to great sensitivity levels
for the first time.  The large survey area (4 square degrees) yields data fit
for such statistical characterizations.

We report our results in this paper.  In the next section
we describe the data and outline our reduction procedure.  In Section
\ref{sec:res} we present our results.  We compare our results with those
in the K-band and discuss their implications in Section \ref{sec:dis}.

\section{Data}
\label{sec:data}

The First Look Survey is designed to provide a characteristic first-look
at the mid-infrared sky to sensitivities that are two orders of magnitudes
deeper than previous large-area surveys.  The Infrared Array Camera (IRAC)
uses four imaging detectors with passbands centered at wavelengths 3.6 $\mu$m,
4.5 $\mu$m, 5.8 $\mu$m, and 8.0 $\mu$m.  The IRAC FLS survey provides a
uniform coverage of a 4 square-degree field centered at $RA=17^{h}18^{m}$,
$Dec=59\arcdeg30\arcmin$ with a total 60-second exposure time for each pixel
in the $256 \times 256$ arrays.

Our starting point is the source catalog for each IRAC channel \citep{lacy04_2}.
In brief the Basic Calibrated Data of each IRAC channel was mosaicked for the
entire FLS field,  source-extracted using SExtractor involving mask and coverage
images, and spurious sources removed.  Aperture corrections were applied to
aperture fluxes in the source lists.

To separate stars from the source lists, we use the FLS R-band catalog
\citep{fadda04} covering the same field.  We
perform a bandmerge with the R-catalog for each source list of the IRAC
channels.  Since no apparent decrease of star counts up to R=24 in the
R-catalog is found \citep{fadda04}, we delete IRAC sources with R-band
stellarity index greater than 0.9 up to R=24.  We verify the
validity of this criteria by establishing histograms of the R-band
stellarity index in several magnitude bins, which show that the stellarity
index peaks at $>0.9$ down to $R>24$ \citep{hat04}.

We band-merge the four IRAC source lists to study the source distribution
in the IRAC color-space before and after the de-contamination of stars
brighter than R=24.  We verify that the star de-contamination significantly
reduces the concentration of sources at colors corresponding to the
Rayleigh-Jeans flux ratios at IRAC wavelengths.  However, the process
also deletes a population of red sources with high R-band stellarity, some
of which may correspond to AGN at high redshifts \citep{lacy04_1}.
We then follow the same color criteria used in \citet{lacy04_1} and
place these sources back to our IRAC source lists.  The number of sources
thus classified as stars and dropped from the samples are 27,049, 21,365,
8,176, and 5,144 for the 3.6$\mu$m, 4.5$\mu$m, 5.8$\mu$m, and 8.0$\mu$m,
respectively.

We estimate the quality of the star/galaxy separation by comparing the
resulting 3.6$\mu$m star and galaxy counts with the existing models. 
A star-counts model from the 2MASS survey \citep{jarrett04} predicts a total
of 27,583 stars to the 5-$\sigma$ sensitivity level in the validated 3.6$\mu$m
area, slightly higher than 27,049 stars from our separation.  The remaining
contamination, mostly by fainter stars, is less than 2.5\%.  We establish
flux-limited samples to avoid selection effects (see below) at 15-$\sigma$.
At these sensitivities the remaining contamination becomes negligible.
We then establish galaxy source counts based on the decontaminated
3.6$\mu$m source list, which fits well to the deep K-band source-counts after
correction based on average galaxy color \citep{wilson04,jarrett04}.
The comparisons of the 3.6$\mu$m source counts with the models show the
general quality of our star/galaxy separation procedures.

To this point the sources in our list are selected in regions of slightly
different IRAC coverages.  To eliminate the bias of source selection across
the field caused by any non-uniform survey coverage, we establish our final
source lists to include sources above 15-$\sigma$ sensitivity levels.
Table 1 lists the flux-limits and number of sources in these source lists.

\placetable{tbl-1}
\placefigure{ang}

\section{Results}
\label{sec:res}

The angular auto-correlation function, $w(\theta)$, measures the excess numbers
of pairs of galaxies in a single catalog as a function of angular separation
compared to those in a random distribution.  We have calculated the angular
correlation function using the \citet{ls93} estimator,

\begin{equation}
\label{eqn:ls}
w({\theta}) = \frac{DD-2DR+RR}{RR}.
\end{equation}

Here DD, DR, and RR are normalized pair-counts in a given angular separation
of the data-data, data-random, and random-random sources, respectively.
For each IRAC-channel data, we generate 10 realizations of
random sources using the mask and coverage files that our source lists are based
on, to ensure similar source selection.  The number of the random points generated
in each realization is approximately identical to the number of galaxies in an IRAC
source list.  

The correlation signal is calculated in each angular separation range after binning
the data.  The uncertainty across different angular bins are not independent but
correlated.  To better estimate the uncertainty in our measurements, we use
the Jackknife analysis \citep{scranton02,maller03} to calculate the uncertainties
for each angular bin by dividing the survey area into 20 tiles.  We calculate
the correlation each time by removing sources in a tile and obtained 20 correlation
estimates based on which the uncertainties of the correlation signal are estimated.
Due to the sample size we perform the Jackknife analysis for the 3.6$\mu$m and
4.5$\mu$m data only.  For the 5.8$\mu$m and 8$\mu$m our sample sizes are not
appropriate for the Jackknife analysis.  We calculated Poisson errors in these
cases $\delta w({\theta}) = \sqrt{(1+w(\theta))/DD}$ \citep{hewitt82}.
However we need to bear in mind that using Poisson uncertainites do not include
the correlation noise between the angular bins and therefore may underestimate
the total noise.
A comparison between the power-law models fits (see below) of using Jackknife
errors and Poisson errors in our 3.6$\mu$m and 4.5$\mu$m data shows that
using Poisson errors can under-estimate the uncertainties in the amplitude
by up to 50\% while over-estimating the amplitude values by up to 22\% in
our cases.

Due to limited sample size the angular correlation calculated this way
are known to be biased to lower values, and the so-called ``Integral constraints''
need to be estimated to correct for the bias.  We calculate the integral constraints
for our samples following \citet{re99}.  The correlation amplitudes quoted
in this paper are all corrected ones.

In Figure \ref{fig:ang} we illustrate the angular correlation functions
we measured at the four IRAC wave-bands.  Comparing with random distributions,
there are strong signals of galaxy clustering in the IRAC wavelengths.  We find that
the four auto-correlation functions are well-approximated by power-laws out to
about 0.2$\arcdeg$.  The amplitude and slope of the correlation vary from
band to band.  We also plot by the solid lines in the Figure the single-power-law model
for each of the measured correlation after corrections for integral constraint
are made.  Table 1 lists the parameters of the models.

\section{Summary and Discussion}
\label{sec:dis}

We have measured the angular 2-point auto-correlation function of galaxies
in the FLS IRAC data.  We find the the angular correlations in IRAC wavelengths
are consistent with power-laws with varying slope and amplitude.

Old stellar populations contribute significantly to IRAC wavelengths.
Galaxies dominated by these populations are mostly elliptical galaxies.
Since these galaxies are known to cluster more strongly \citep{dress80}
than other types of galaxies where younger stellar populations dominate,
we would expect to detect strong clustering in IRAC surveys.  Our estimate
of the correlation amplitude is higher than but marginally consistent with
what is measured in the 3.6$\mu$m channel in the SWIRE Lockman and ELAIS N1
fields \citep{oliver04}, where a slightly steeper correlation slope is
found. 

For IRAC channels, especially at 3.6$\mu$m, the nearby K-band has extensive
measurements of galaxy angular correlations to various depths.  In Figure
\ref{fig:ampflux} we compare our full-sample measurement with various
results from the K-band magnitude-limited samples in the recent literature.
In order to make these comparisons, we use a fixed slope, $\gamma=1.8$,
to re-calculate the integral constraint and the correlation amplitude in
our 3.6$\mu$m sample limited at 15-$\sigma$.  The same fixed-slope has
been used in most of the K-band measurements we compare with.  We obtain
an amplitude of $(6.81\pm1.00)\times10^{-4}$.  We use an average color
of 1.1 \citep{oliver04} to scale the K-magnitude limits in various
measurements to the 3.6$\mu$m magnitude and flux.  The Figure also shows
the uncertainty of this color-scaling.  Based on the Figure we conclude that
our measurement is only marginally consistent with the K-band results
from the various magnitude-limited samples.  Our lower estimate of the
correlation amplitude in the 3.6$\mu$m sample may indicate a different
galaxy clustering property in this wavelength.  However the difference
is only marginal given the uncertainties in the color scaling.
Since these wavelengths detect the same population of galaxies,
and the rest-frame K-band sources also become IRAC detections at higher
redshifts, there is good reason to believe that the galaxy clustering
strengths in these wavelengths are close.  Better SED models can resolve
this as more IRAC data becomes available.

The galaxy angular correlation is a result of evolving spatial correlation
scaled by a radial selection function integrated over redshifts.  The
radial selection function determines how galaxies are selected by a survey.
They are formally related by the Limber equation \citep{limber53}.
Using a form of K-band selection function \citep{baugh96} we perform
the Limber integration of the 3.6$\mu$m measurement by assuming a spatial
correlation amplitude of $r_{0}=4.2h^{-1}$Mpc and a slope of 1.8 \citep{oliver04}.
In Figure \ref{fig:limber} we plot the results.  This places constraints on
the median redshift of the sample we used.  Depending on evolution models
for galaxy clustering, the median redshift of the 3.6$\mu$m sample is
$\sim0.42$ if clustering is fixed in proper coordinates, $\sim0.60$ if
clustering is fixed in comoving coordinates.  Bracketed by these two evolution
models, the exact value of the median redshift of the sample would be
within this range if the selection function applies to the sample.  However,
so far we do not have enough information to determine if the form of the
selection function used is appropriate, therefore the uncertainties in
the median redshift estimate may be significant.  The exact value shall be
determined when redshifts of at least a sample representative of the
distribution of sources in both depth and type are measured, at which point
a reliable form of the spatial correlation and its evolution can be constrained.

To investigate the change of correlation function with flux, we establish
two sub-samples at different flux limits for each IRAC channel and calculate
the correlation function in these subsamples.  The results are listed in
Table 1.  In all four IRAC passbands the correlation amplitudes decrease
with increasing flux-limit values in the established subsamples.  If we
again assume the form of the selection function and the parameters of
spatial correlations used in Figure \ref{fig:limber}, and that clustering
is fixed in proper coordinates, we estimate that the median redshifts are
0.25 and 0.13, respectively for the 3.6$\mu$m ``sub1'' and ``sub2'' in Table 1.
While we believe that the change of correlation amplitudes in subsamples of
all IRAC channels has cosmological origin, we caution that the scaling
of these to median redshifts using Figure \ref{fig:limber} may lack accuracy,
based on the above discussion.  The results can be more fully interpreted
when we have more complete information about the redshift distribution of
the sources in these samples.  We plan to pursue this as part of our
future work.

\acknowledgments 

We wish to thank an anonymous referee for help shaping up the final form
of this paper.  This work is based on observations made with the Spitzer
Space Telescope, which is operated by the Jet Propulsion Laboratory (JPL),
California Institute of Technology under NASA contract 1407.  Support for
this work was provided by NASA through JPL.



\clearpage

\figcaption[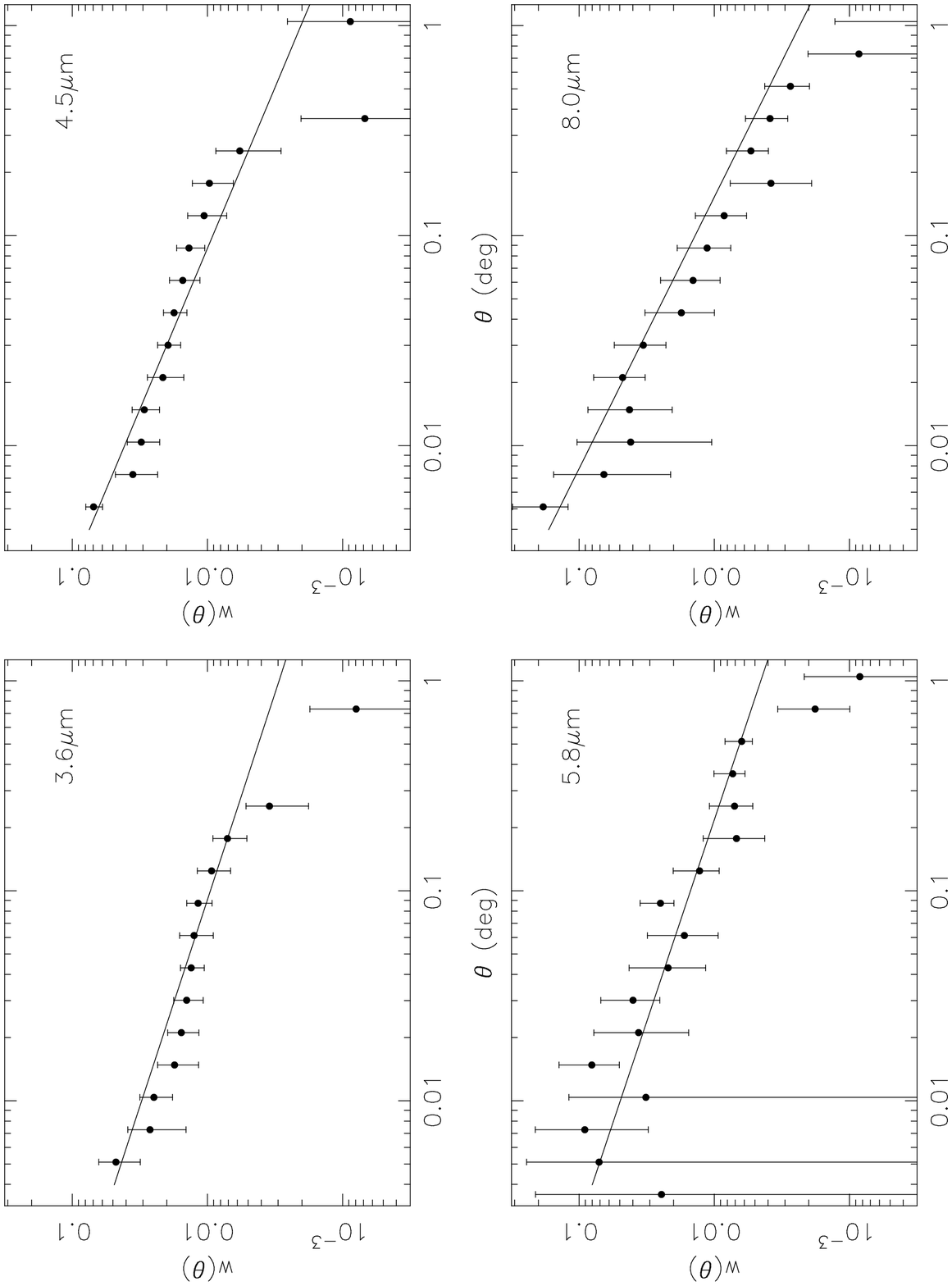]{
The angular 2-point auto-correlation functions measured for the IRAC channels.
A power-law model $w(\theta)=A\theta^{1-\gamma}$ is shown by the lines, where
corrections for the integral constraint are included.  The Jackknife analysis
is performed for the 3.6$\mu$m and 4.5$\mu$m, represented by the error bars.
The uncertainties for 5.8$\mu$m and 8.0$\mu$m are Possion error which may
underestimate the total error.
\label{fig:ang}}

\figcaption[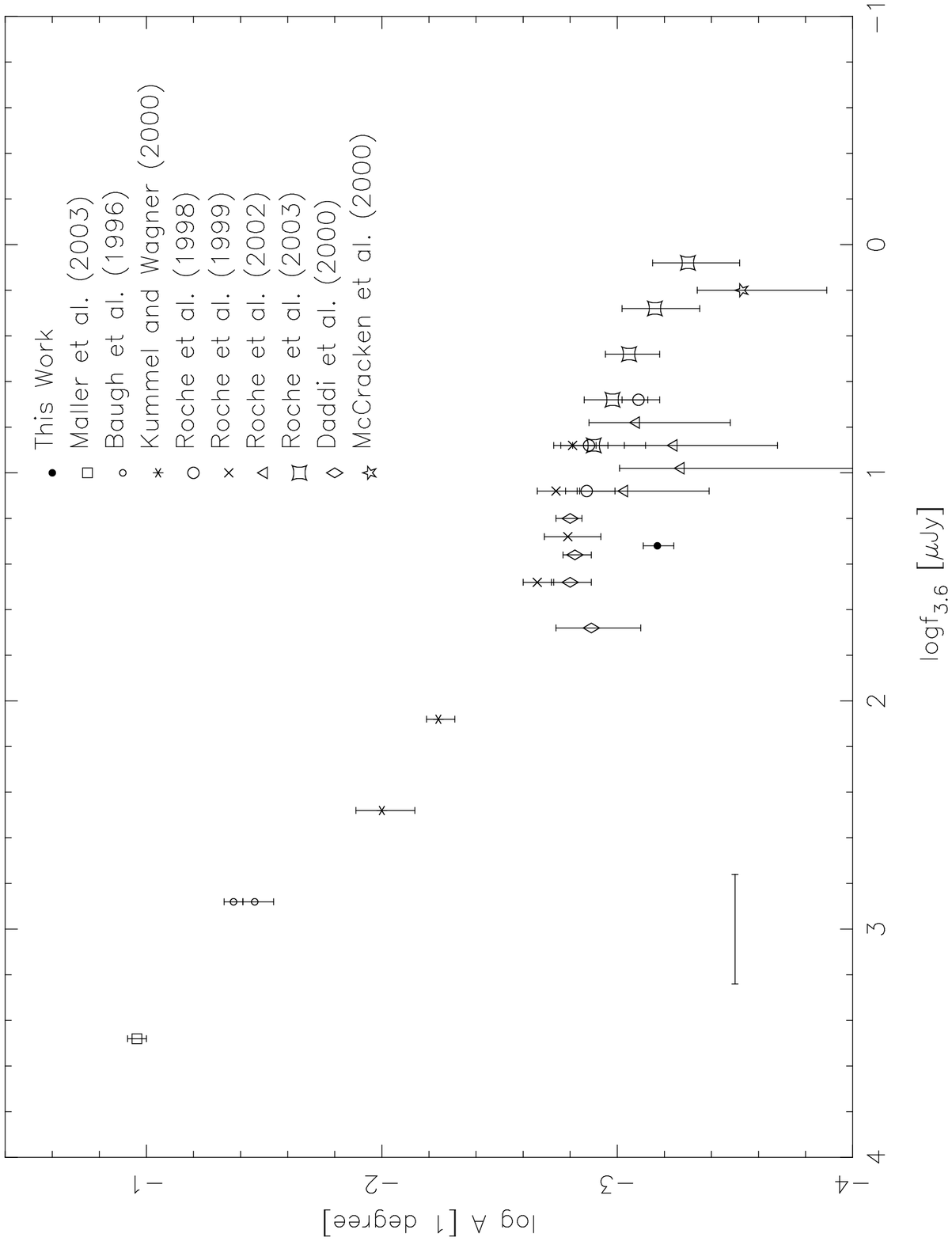]{
Our fixed-slope ($\gamma=1.8$) estimate of the angular correlation amplitude
at 3.6$\mu$m as compared with the K-band results from the literature.  We scale
the K-band magnitudes to the 3.6$\mu$m fluxes using an average color of 1.1. 
The horizontal error bar indicates the uncertainty of the color scaling for
the K-band points.  The fluxes represent the flux limit of a given sample.
To be consistent, we only plot the result from the 15-$\sigma$ flux-limited
sample at 3.6$\mu$m.
\label{fig:ampflux}}

\figcaption[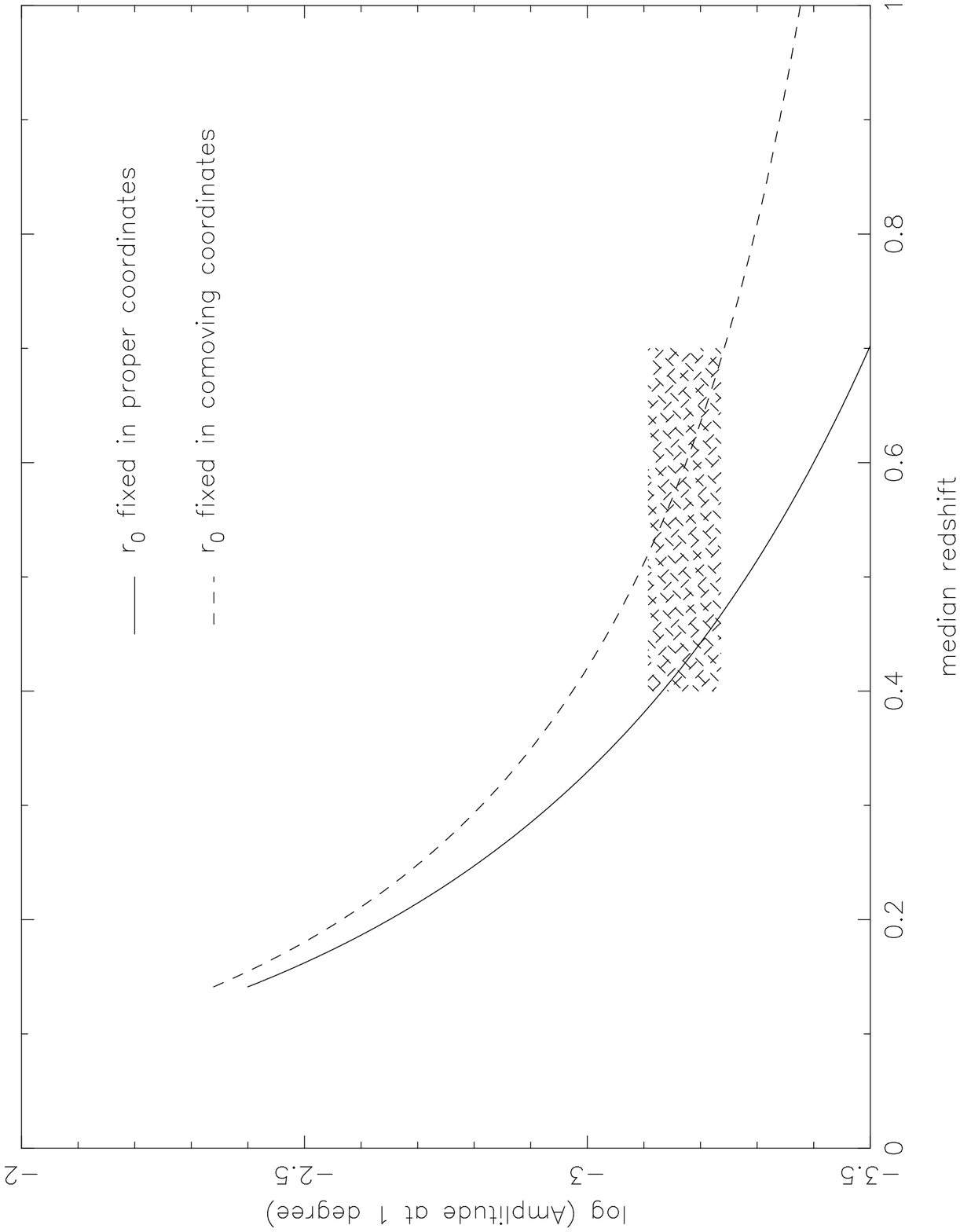]{
Change of the angular correlation amplitude with the sample median redshift
based on Limber integration.  Here we use a form of selection function
\citep{baugh96} that has been studied in K-band surveys, and assumed a spatial
correlation amplitude estimated from a 3.6$\mu$m SWIRE sample.  The two evolution
models for galaxy clustering along with the uncertainties from our angular
correlation amplitude measurement places a contraint on the median redshift of
the 3.6$\mu$m sample in the FLS survey.
\label{fig:limber}}

\newpage
\plotone{f1.eps}
\newpage
\plotone{f2.eps}
\newpage
\plotone{f3.eps}

\clearpage

\begin{deluxetable}{lccccc}
\tablewidth{0pt}
\tablecaption{IRAC Samples and Angular Correlation functions\tablenotemark{1} \label{tab:ang}}
\tablehead{
\colhead{Sample}    & \colhead{flux limit}        & \colhead{Galaxy Number} & \colhead{A ($10^{-3}$)} & \colhead{$\gamma$} & \colhead{IC}}
\startdata
3.6{$\mu$m} total   & $f_{\nu}>$18$\mu$Jy           & 52790                   & 2.95$\pm$1.84           & 1.51$\pm$0.32     & 1.14 \\
3.6{$\mu$m} sub1    & 48$\mu$Jy$<f_{\nu}<$78$\mu$Jy   & 9101                    & 2.83$\pm$0.44           & 1.65$\pm$0.30   & 1.22 \\
3.6{$\mu$m} sub2    & 78$\mu$Jy$<f_{\nu}<$120$\mu$Jy  & 4354                    & 4.33$\pm$0.11           & 1.79$\pm$0.29   & 1.31 \\
4.5{$\mu$m} total   & $f_{\nu}>$21$\mu$Jy           & 41936                   & 2.03$\pm$0.45           & 1.65$\pm$0.30     & 1.22 \\
4.5{$\mu$m} sub1    & 56$\mu$Jy$<f_{\nu}<$91$\mu$Jy   & 5615                    & 1.95$\pm$0.06           & 1.85$\pm$0.27   & 1.36 \\
4.5{$\mu$m} sub2    & 91$\mu$Jy$<f_{\nu}<$140$\mu$Jy  & 2498                    & 9.87$\pm$1.10           & 1.56$\pm$0.29   & 1.17 \\
5.8{$\mu$m} total   & $f_{\nu}>$135$\mu$Jy          & 4207                    & 4.53$\pm$1.10           & 1.52$\pm$0.26     & 1.15 \\
5.8{$\mu$m} sub1    & 135$\mu$Jy$<f_{\nu}<$180$\mu$Jy & 1118                    & 7.27$\pm$0.18           & 1.69$\pm$0.26   & 1.24 \\
5.8{$\mu$m} sub2    & 180$\mu$Jy$<f_{\nu}<$270$\mu$Jy & 1108                    & 11.40$\pm$0.30           & 1.64$\pm$0.26  & 1.21 \\
8.0{$\mu$m} total   & $f_{\nu}>$96$\mu$Jy           & 5721                    & 2.34$\pm$0.13           & 1.77$\pm$0.28     & 1.30 \\
8.0{$\mu$m} sub1    & 96$\mu$Jy$<f_{\nu}<$128$\mu$Jy  & 1491                    & 5.77$\pm$0.13           & 1.72$\pm$0.25   & 1.26 \\
8.0{$\mu$m} sub2    & 128$\mu$Jy$<f_{\nu}<$192$\mu$Jy & 1577                    & 8.23$\pm$0.30           & 1.64$\pm$0.25   & 1.21 \\
\enddata
\tablenotetext{1}{A power-law form $w(\theta)=A(\theta^{1-\gamma}-IC)$ is used to
fit the data where IC is the integral constraint; $\theta$ is in degrees.}
\end{deluxetable}


\begin{thebibliography}{}

\bibitem[Baugh et al. 1996]{baugh96} Baugh, C. M., Gardner, J. P., Frenk, C. S., \& Sharples, R. M. 1996, \mnras, 283, L15.

\bibitem[Daddi et al. 2000]{daddi00} Daddi, E. et al. 2000, \aap, 361, 535. 

\bibitem[Fadda et al. 2004]{fadda04} Fadda, D., Jannuzi, B., Ford, A., \& Storrie-Lombardi, L. J. 2004, \aj, accepted.

\bibitem[Groth \& Peebles 1977]{gp77} Groth, E. J. \& Peebles, P. J. E. 1977, \apj, 217, 385.

\bibitem[Dressler 1980]{dress80} Dressler, A. 1980, \apj, 236, 351.

\bibitem[Hatziminaoglou 2004]{hat04} Hatziminaoglou, E. 2004, private communication.

\bibitem[Hewitt 1982]{hewitt82} Hewitt, P. C. 1982, \mnras, 201, 867.

\bibitem[Jarrett 2004]{jarrett04} Jarrett, T. 2004, in preparation.

\bibitem[Kummel \& Wagner 2000]{kw00} Kummel, M. W. \& Wagner, S. J. 2000, \aap, 353, 867.

\bibitem[Lacy et al. 2004]{lacy04_1} Lacy, M. et al. 2004, this issue.

\bibitem[Lacy et al. 2004]{lacy04_2} Lacy, M. et al. 2004, in preparation.

\bibitem[Limber 1953]{limber53} Limber, D. N., 1953, \apj, 117, 134.

\bibitem[Landy \& Szalay 1993]{ls93} Landy, S. D. \& Szalay, A. S. 1993, \apj, 412, 64.

\bibitem[Maller et al. 2003]{maller03} Maller, A. H., McIntosh, D. H., Katz, N., \& Weinberg, M. D. 2003, astro-ph/0304005.

\bibitem[McCracken et al. 2000]{mc00} McCracken, H. J., Shanks, T., Metcalfe, N., Fong, R., \& Campos, A. 2000, \mnras, 318, 913.

\bibitem[Oliver et al. 2004]{oliver04} Oliver, S. et al. 2004, \apj, this issue.

\bibitem[Roche et al. 1998]{roche98} Roche, N., Eales, S., \& Hippelein, H. 1998, \mnras, 295, 946.

\bibitem[Roche \& Eales 1999]{re99} Roche, N. \& Eales, S. A. 1999, \mnras, 307, 703.

\bibitem[Roche et al. 1999]{roche99} Roche, N., Eales, S. A., Hippelein, H., \& Willott, C. J. 1999, \mnras, 306, 538.

\bibitem[Roche et al. 2002]{roche02} Roche, N., Almaini, O., Dunlop, J., Ivison, \& Willott, C. J. 2002, \mnras, 337, 1282.

\bibitem[Roche et al. 2003]{roche03} Roche, N., Dunlop, J. \& Almaini, O. 2003, \mnras, 346, 803.

\bibitem[Scranton et al. 2002]{scranton02} Scranton, R. et al. 2002, \apj, 579, 48.

\bibitem[Wilson et al. 2004]{wilson04} Wilson, G. et al. 2004, in preparation.

\end{thebibliography}
\end{document}